\documentclass[10pt, conference]{IEEEtran}
\IEEEoverridecommandlockouts
\usepackage{cite}
\usepackage{amsmath,amssymb,amsfonts}
\usepackage{algorithmic}
\usepackage{graphicx}
\usepackage{textcomp}
\usepackage{xcolor}
\def\BibTeX{{\rm B\kern-.05em{\sc i\kern-.025em b}\kern-.08em
    T\kern-.1667em\lower.7ex\hbox{E}\kern-.125emX}}

\usepackage{physics}
\usepackage{yfonts}
\usepackage[export]{adjustbox}
\usepackage[inline, shortlabels]{enumitem}
\usepackage{subcaption}
\usepackage{bm}
\usepackage{hyperref}
\newcommand{\figref}[1]{Fig.~\ref{#1}}
\newcommand{\tabref}[1]{Tab.~\ref{#1}}
\newcommand{\secref}[1]{Section~\ref{#1}}
\newcommand{\etal}{\textit{et al.}}

\begin{document}

\title{Approximately Equivariant Quantum Neural Network for $p4m$ Group Symmetries in Images
}


\author{\IEEEauthorblockN{Su Yeon Chang\IEEEauthorrefmark{1}\IEEEauthorrefmark{2}, Michele Grossi\IEEEauthorrefmark{1}, Bertrand Le Saux\IEEEauthorrefmark{3} and Sofia Vallecorsa\IEEEauthorrefmark{1}
}\IEEEauthorblockA{\IEEEauthorrefmark{1}\textit{IT Department}, \textit{European Organization for Nuclear Research (CERN)}, \textit{CH-1211 Geneva, Switzerland}\\
\IEEEauthorrefmark{2} \textit{Laboratory of Theoretical Physics of Nanosystems (LTPN), Institute of Physics,} \\ 
\textit{\'Ecole Polytechnique F\'ed\'erale de Lausanne, CH-1015 Lausanne, Switzerland}}
\IEEEauthorblockA{\IEEEauthorrefmark{3}\textit{$\Phi$-lab} \textit{European Space Agency, IT-00044, Italy}}
Email: \href{mailto:su.yeon.chang@cern.ch}{su.yeon.chang@cern.ch}
}
\maketitle

\begin{abstract}
Quantum Neural Networks (QNNs) are suggested as one of the quantum algorithms which can be efficiently simulated with a low depth on near-term quantum hardware in the presence of noises. However, their performance highly relies on choosing the most suitable architecture of Variational Quantum Algorithms (VQAs), and the problem-agnostic models often suffer issues regarding trainability and generalization power. As a solution, the most recent works explore Geometric Quantum Machine Learning (GQML) using QNNs equivariant with respect to the underlying symmetry of the dataset. GQML adds an inductive bias to the model by incorporating the prior knowledge on the given dataset and leads to enhancing the optimization performance while constraining the search space. This work proposes equivariant Quantum Convolutional Neural Networks (EquivQCNNs) for image classification under planar $p4m$ symmetry, including reflectional and $90^\circ$ rotational symmetry. We present the results tested in different use cases, such as phase detection of the 2D Ising model and classification of the extended MNIST dataset, and compare them with those obtained with the non-equivariant model, proving that the equivariance fosters better generalization of the model.   
\end{abstract}


\section{\label{sec:intro}Introduction}
During the last few years, Quantum Machine Learning (QML) has witnessed remarkable progress from diverse research perspectives as a promising application for the practical use of quantum computers~\cite{Schuld2021QML, Havlek2019, Cerezo2021VQA}. In particular, Quantum Neural Networks (QNNs) are suggested as the most general and fundamental formalism to solve a plethora of problems while the architecture is completely agnostic to the given problem. They are often expected to improve existing machine learning (ML) techniques in terms of training performance~\cite{Havlek2019, Huang2021PowerOfData}, convergence rate~\cite{Liu2021Speedup, Llyod2013Speedup}, and generalization power~\cite{Caro2022generalization, Monaco2023Generalization, siemaszko2022rapid}. 

However, they often suffer from issues due to complex loss landscapes, which are often non-convex and lead to many poor local minima~\cite{Holmes2022BP, Pesah2021BP, Cerezo2021bp}.  In an effort to solve these issues, the field of Geometric Quantum Machine Learning (GQML)~\cite{kazi2023GQML, Meyer2023GQML, zheng2022GQML, Ragone2023GQML, Larocca2022GQML, schatzki2022GQML, nguyen2022theory} is currently emerging, inspired by the classical Geometric Deep Learning (GDL)~\cite{bronstein2021GDL, Cohen2016group, Batzner2022Equiv}. 

The main idea of GQML is to add sharp inductive bias~\cite{jonas2021bias} into the training model by incorporating prior knowledge on the dataset~\cite{nguyen2022theory, schatzki2022GQML}. In practice, GQML aims to construct a parameterized QNN, which is equivariant under the action of the symmetry group associated with the input dataset, so that the same action is applied to the output of the QNN. Previous studies have heavily explored GQML both in terms of theories and applications, showing that GQML helps mitigate the issues often encountered in QML~\cite{nguyen2022theory}. However, most studies still focus on the permutation symmetric group, $S_n$~\cite{kazi2023GQML, schatzki2022GQML, zheng2022GQML}, $\mathbb{Z}_2\otimes\mathbb{Z}_2$ symmetry applied in small toy applications~\cite{Meyer2023GQML}, or a single symmetry element in the case of image classification~\cite{West2023Reflection}. 

We extend the study on GQML in the context of image classification by taking into account the \textit{planar wallpaper symmetry} group, $p4m$, which includes the reflection and the $90^\circ$ rotation. The $p4m$ symmetry group is the most common symmetry group observed in image datasets, which are already treated in classical GDL via Group Equivariant Convolutional Networks (GCNN)~\cite{Cohen2016group, Dave2021group}. Although symmetry in images is considered to hinder neural network training in general, there exist applications where the symmetry has ponderable importance, such as Earth Observation~\cite{Marcos_2017_ICCV, Paoletti2020EO, Sebastianelli2022, Chang2022 }, medical images~\cite{Yan2020Medical}, symmetry-related physics datasets~\cite{Bogatskiy2020HEP}, etc. 

In this work, we introduce the $p4m$-\textit{Equivariant Quantum Convolutional Neural Network} (EquivQCNN) for image classification. The results clearly prove that the equivariant neural network has the advantage in terms of generalization power, in particular, while using only a small number of training samples. Moreover, we show that the presence of small noise in the EquivQCNN training helps to classify the symmetric images better. Our study ultimately paves the way for GQML to tackle realistic image classification tasks, improving training performance. 

This paper is structured as follows. First of all, we will briefly summarize in \secref{sec:preliminaries} the theoretical backgrounds required to understand GQML. Then, in \secref{sec:eqcnn}, we introduce the architecture of \textit{Equivariant Quantum Convolutional Neural Network} (EquivQCNN) for the \textit{planar wallpaper symmetry} group $p4m$ in the context of image classification. \secref{sec:result} present our first result for EquivQCNN applied on reflectional and rotational symmetric images and prove its generalization power compared to the non-equivariant architecture. We finally conclude in \secref{sec:conclusion} and propose a future research direction. 

\section{\label{sec:preliminaries}Preliminaries}
This section summarizes the theoretical background on group symmetry and equivariance required to construct a GQML architecture for supervised learning. Consider a classical data space $\mathcal{X}$ and a label space $\mathcal{Y}$. Each data sample $\mathbf{x}\in\mathcal{X}$ is associated with a label $\ell \in \mathcal{Y}$ with the underlying function $f:\mathcal{X}\to \mathcal{Y}$.  The supervised learning aims to find $y_{\bm{\theta}}$, which is as close as possible to the ground truth $f$ with the trained parameters $\bm{\theta}$.  

In the case of QML, we construct a quantum feature map $\psi:\mathcal{X}\to\mathcal{H}$, which embeds the classical data into a quantum state in the Hilbert space $\mathcal{H}$. 
The input quantum state $\ket{\psi(\mathbf{x})}\in\mathcal{H}$ is then transformed via QNN, taking the most general form of a Variational Quantum Circuits (VQCs) $\mathcal{U}(\bm{\theta})$ which is parameterized by the rotation angles $\bm{\theta}$. The final prediction of the QNN for the input feature $\mathbf{x}$ is given as an expectation value of the observable $O$ : 
\begin{equation}
    y_{\bm{\theta}}(\mathbf{x}) = \bra{\psi(\mathbf{x})}\mathcal{U}^\dagger(\bm{\theta})O\mathcal{U}(\bm{\theta})\ket{\psi(\mathbf{x})}. 
\end{equation} 

In general, QNN architecture is completely agnostic to the underlying symmetry of $\mathcal{X}$. GQML aims to incorporate the symmetry inherent to the dataset with the QNN architecture so that the final prediction is \textit{invariant} after the action of the symmetry group element on the original input feature.

\begin{figure*}
    \centering
    \includegraphics[width = \textwidth, trim = {0.3cm, 5.2cm, 2.5cm, 9.3cm}]{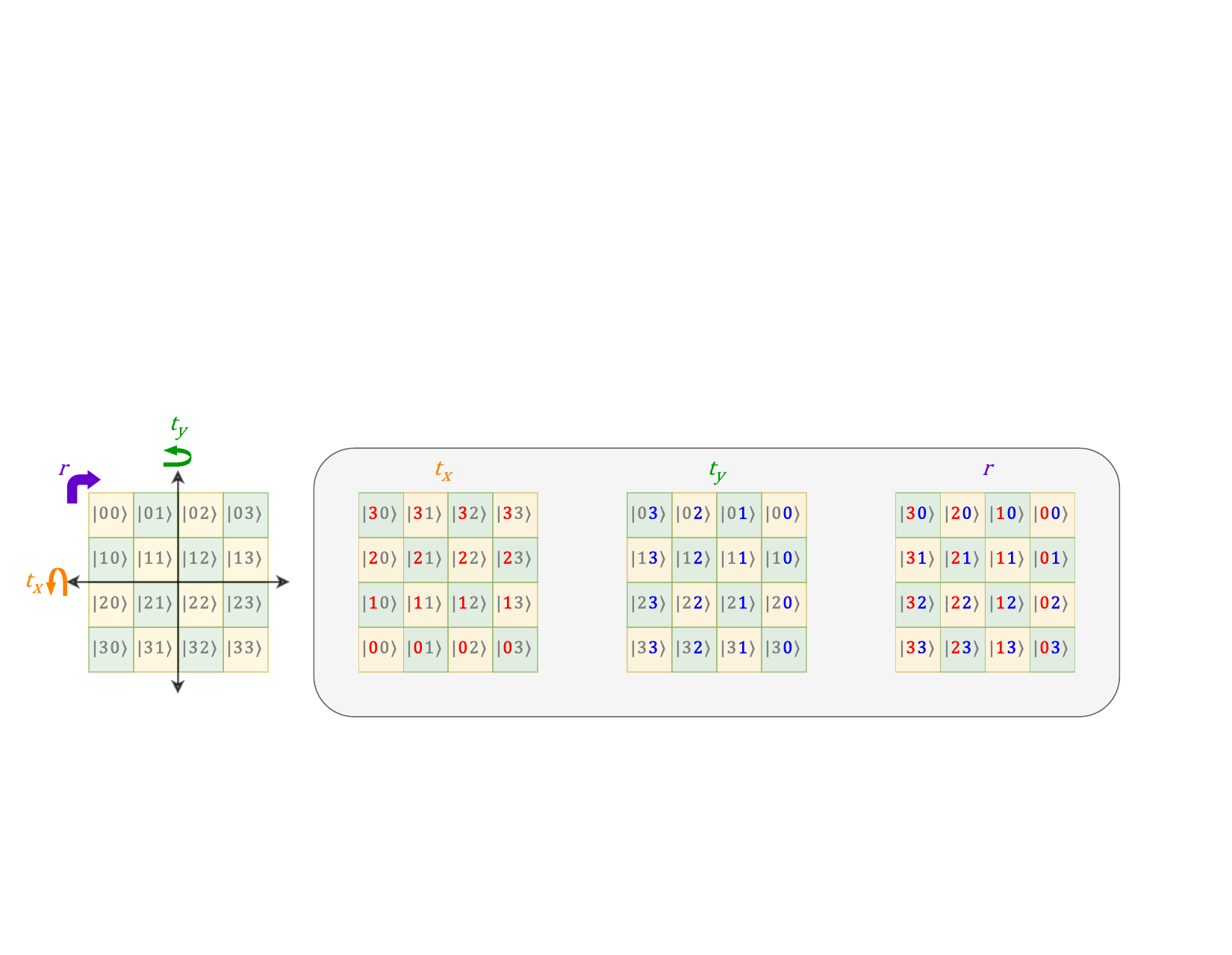}
    \caption{Schematic diagram of the action of $p4m$ symmetry on 2D images of size $4 \times 4$ encoded using CAA embedding method  with $2$ qubits. The pixel at position $(i,j)$  is associated with a computational basis $\ket{i}\ket{j}$.  }
    \label{fig:symmetry}
\end{figure*}

Let us formalize it in a more concrete way. Consider a symmetry group $\frak{G}$ that acts on the data space $\mathcal{X}$. We say the training model is $\frak{G}$\textit{-invariant} if : 
\begin{equation}
    y_{\bm{\theta}}(g[\mathbf{x}]) = y_{\bm{\theta}}(\mathbf{x}), \forall \mathbf{x} \in \mathcal{X}, \forall g \in \frak{G}.
\end{equation}

In order to construct $\frak{G}$\textit{-invariant} model, we require three components: equivariant data embedding, equivariant QNN and invariant measurement~\cite{Meyer2023GQML}. 

First of all, we say that the data embedding is $\frak{G}$-\textit{equivariant} if the symmetry group element $g\in\frak{G}$ applied on the data $\mathbf{x}\in\mathcal{R}$ induces a unitary quantum action $V_s[g]$ in the level of quantum states :  
\begin{equation}
    \ket{\psi{(g[\mathbf{x}])}} = V_s[g] \ket{\psi{(\mathbf{x})}}.
\end{equation}
We call $V_s$ the \textit{induced representation} of the embedding $\psi(\mathbf{x})$~\cite{Meyer2023GQML}. 

We also need to construct a trainable quantum circuit ansatz, parameterized by angles $\bm{\theta}$, which is equivariant with respect to the symmetry group $\frak{G}$. 
For simplicity, we consider only the gates generated by a fixed generator $G\in \mathcal{G}$ : 
\begin{equation}
    R_G(\theta) = \exp(-i\theta G),~\theta\in\mathbb{R}. 
\end{equation}
where $\mathcal{G}$ is a fixed gateset. 
For a symmetry group $\frak{G}$ and its representation $V_s$, the operator $R_G$ is said to be $\frak{G}$-equivariant if and only if~\cite{Larocca2022GQML, Meyer2023GQML}: 
\begin{equation}
    [R_G(\theta), V_s[g]] = 0, \forall g \in \frak{G}, \forall \theta \in \mathbb{R} 
\end{equation}
or equivalently, 
\begin{equation}
    [G, V_s[g]] = 0, \forall g \in \frak{G}. 
\end{equation}
The definition of \textit{equivariance} can also be extended to QNNs. We call that a QNN, $\mathcal{U}_{\theta}$, is $\frak{G}$-\textit{equivariant} if and only if it consists of equivariant quantum operators, i.e. $\mathcal{U}_{\theta}$ commutes with all the components in the symmetric group $\frak{G}$. 

There exist several methods to construct the equivariant gateset~\cite{nguyen2022theory}, but in this paper, we will focus on the \textit{twirling method}, which is the most practical approach for small symmetry groups~\cite{Helsen2019Twirling}. Consider an arbitrary generator $X$. Then, we define a twirled operator $\mathcal{T}_{\frak{G}}$ as : 
\begin{equation}
    \mathcal{T}_{\frak{G}}[X] = \frac{1}{|\frak{G}|}\sum_{g \in \frak{G}} V_s[g]^\dagger X V_s[g]. 
\end{equation}
It corresponds to a projector of the operator $X$ onto all symmetry group elements, commuting with $V_s[g]$ for all $g\in \frak{G}$. 

Finally, an observable $O$ is $\frak{G}$-invariant, if : 
\begin{equation}
    V_s[g]^\dagger O V_s[g] = O,~\forall g \in \frak{G}, 
\end{equation}
i.e. if $O$ commutes with $V_s[g] $ for all $g \in \frak{G}$. By taking all three components,  the equivariance of QNN leads to the invariance of the final prediction:
\begin{align}
    y(g[\mathbf{x}]) & = \bra{\psi(g[\mathbf{x}])}\mathcal{U}(\bm{\theta})^\dagger O\mathcal{U}(\bm{\theta}) \ket{\psi(g[\mathbf{x}])} \nonumber \\ & =   \bra{\psi(\mathbf{x})}V_s^\dagger \mathcal{U}(\bm{\theta})^\dagger O\mathcal{U}(\bm{\theta})  V_s\ket{\psi(\mathbf{x})} \nonumber \\ 
    & =  \bra{\psi(\mathbf{x})} \mathcal{U}(\bm{\theta})^\dagger O\mathcal{U}(\bm{\theta})  \ket{\psi(\mathbf{x})}  = y(\mathbf{x}) 
\end{align}

Overall, we have a trade-off between the equivariance and the expressibility of the QNN by constraining the quantum operators in the model based on the geometric prior of the dataset. GQML reduces the search space for training and brings advantages in many folds, such as trainability~\cite{schatzki2022GQML}, convergence rate~\cite{Zheng2023Speedup}, and generalization power~\cite{Meyer2023GQML,schatzki2022GQML}.

\section{\label{sec:eqcnn}GQML for image classification}
In this section, we will introduce Equivariant Quantum Convolutional Neural Networks (EquivQCNNs) for image classification invariant under the $p4m$ wallpaper symmetry group, $\frak{G}_{p4m}$ , which corresponds to the planar square symmetry group. It consists of 8 components : 
\begin{itemize}
    \item the identity $e$,
    \item the rotation $r$, $r^2$ and $r^3$ of $90^\circ, 180^\circ, 270^\circ$ around the origin,
    \item the reflection $t_x$ and $t_y$ in the $x$ and $y$ axis,
    \item the reflection in the two diagonals. 
\end{itemize}
In this paper, we will focus on six components out of them, the rotation $r$ and the reflection in the main axis, $t_x$ and $t_y$. 

\subsection{Equivariant Data embedding}

We will start by constructing the data embedding method for the reflectional and rotational symmetry of images. 
Amplitude encoding is one of the most fundamental methods for mapping classical data into quantum states~\cite{Schuld2019Feature}. In general, each pixel coordinate is associated with a computational basis by visualizing the 2D image as a 1D vector, but this complicates the manipulation of 2D symmetry. 

We propose a \textit{coordinate-aware} amplitude (CAA) embedding method, which facilitates finding the unitary representation of $p4m$ symmetry group.  
The main idea of the CAA embedding is that we can explicitly denote the $x$ and $y$ coordinates by using the first $n$ qubits to represent the $x$-coordinate and the second $n$ qubits for the $y$-coordinate of the pixel. 

Let us consider a training set $\mathcal{X}$ of 2-dimensional images with $N \times N$ pixels, denoted as $\mathbf{x} = \{x_{ij}\}$ with $i, j = 0,..., N -1$, each of which is associated with a hot encoded label $\ell \in \{0,1\}  \in \mathcal{Y} $. The CAA embedding maps the input image $\mathbf{x}$ into a quantum state $ \ket{\psi(x)} \in \mathcal{H}$ as follows: 
\begin{equation}
  \ket{\psi(x)} = \sum_{i=0}^{N-1}\sum_{j=0}^{N-1} x_{ij} \ket{i}\ket{j}
\end{equation}
where $N = 2^n$.
For simplicity, let us denote $q_{1:n}$ the first $n$-qubits for $x$-coordinates and $q_{n+1:2n}$ the second $n$-qubits for $y$-coordinates. 
\begin{figure*}[h]
\centering
\begin{subfigure}{0.45\textwidth}
\centering
    \includegraphics[width = \textwidth, trim = {0cm, 0.8cm, 0cm, 2.4cm}]{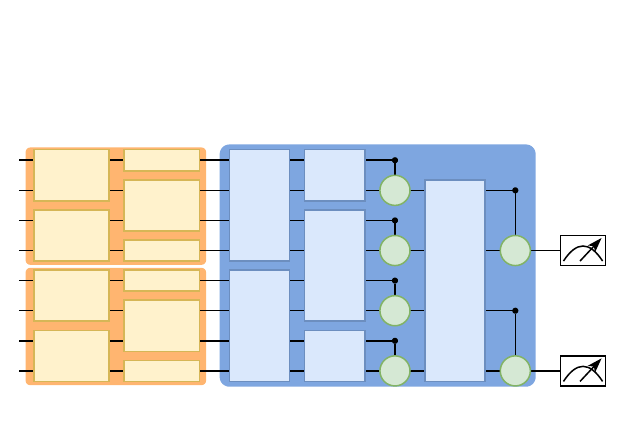}
    \caption{EquivQCNN}
    \label{fig:my_label}
\end{subfigure}
\begin{subfigure}{0.45\textwidth}
\centering
    \includegraphics[width = \textwidth, trim = {0cm, 1.8cm, 0cm, 1.4cm}]{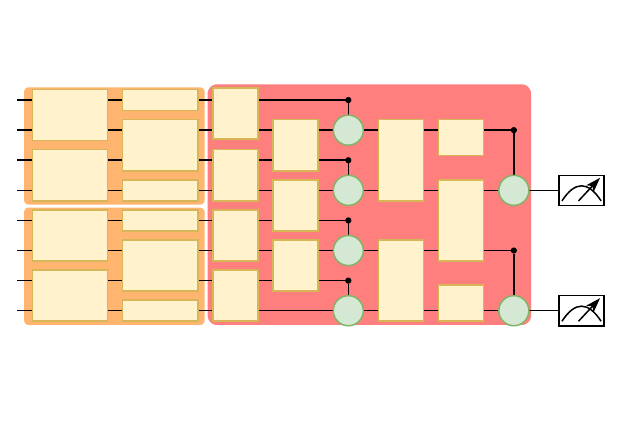}
    \caption{Appr-EquivQCNN}
    \label{fig:my_label}
\end{subfigure}
\caption{A Schematic diagram of (a) EquivQCNN and (b) Appr-EquivQCNN for an example of 8 qubits to classify image of size $16\times 16$. They consist of $U_2$ (yellow rectangle) and $U_4$ (blue rectangle) convolutional filters (c.f. \figref{fig:circuit}), followed by pooling layers (green circle). Both models contain a preliminary \textit{scanning} phase, where $U_2$ acts on $q_{1:n}$ and $q_{n+1:2n}$ separately. EquivQCNN then consists of $U_4$ ansatz, while Appr-EquivQCNN is subject to a small noise by connecting $q_{1:n}$ and $q_{n+1:2n}$ with $U_2$ gate.  }
\label{fig:QCNN}
\end{figure*}

From the CAA embedding formulation, it is straightforward to find the induced representation of $p4m$ group elements. \figref{fig:symmetry} visualizes CAA embedding of 2D images and the action of the symmetry elements on the computational basis states. 
Let us denote $V_{x}$ and $V_{y}$ the induced representation of the reflections, $t_x$ and $t_y$ respectively, and $V_{r}$ for rotation of $90^\circ$, $r$, which are defined as follows : 
\begin{equation}
    V_{x}  = X^{\otimes n} \otimes \mathbb{I}^{\otimes n} = X_{1:n},
\end{equation}
\begin{equation}
    V_y = \mathbb{I}^{\otimes n} \otimes X^{\otimes n} = X_{n+1:2n},
\end{equation}
\begin{equation}
    V_{r}  = (X^{\otimes n} \otimes\mathbb{I}^{\otimes n}) \otimes_{i=0}^{n-1}SWAP_{i, i+n}   = V_x V'_r,
\end{equation}
with $V'r = \otimes_{i=0}^{n-1}SWAP_{i, i+n}$
Therefore, the quantum gates, which are equivariant with respect to $p4m$ symmetry, should commute with all the induced representation, $V_{p4m} = \{V_x, V_y, V_{r}\}$ :  
\begin{align}
    U & \in \frak{comm}\{V_x, V_y, V_{r} \} = \frak{comm}\{V_x, V_y, V'_{r} \}\nonumber \\ 
    & = \frak{comm}\{X_{1:n}, X_{n:2n}, \otimes_{i=0}^{n-1}SWAP_{i, i+n}\}, 
\end{align}
where $\frak{comm}$ denotes the commutator of the unitary operators. 

\subsection{Equivariant Quantum Convolutional Neural Networks }
First proposed by Iris Cong, \etal ~\cite{Cong2019QCNN}, Quantum Convolutional Neural Networks (QCNNs) is the quantum analogue of classical Convolutional Neural Networks (CNNs).
QCNNs have exhibited success in different tasks, including quantum many-body problems~\cite{Cong2019QCNN}, phase detection~\cite{Monaco2023Generalization}, and image classification~\cite{Hur2022QCNN}, taking advantage of avoiding the barren plateaus with shallow circuit depth~\cite{Pesah2021BP}. 

QCNN consists of two components, \textit{convolutional filters}, which are $k$-body local quantum gates for $k < n$, and the pooling layers, to reduce the two qubit states into one qubit state. In most of the cases, we have $k=2$ for convolutional filters, but in this paper, we will also introduce the case with $k>2$ to maintain the equivariance. 
Especially QCNN maintains the translational invariance of input data by sharing identical parameters between the filters inside each layer.  

Following the definition of equivariance and the method presented in \secref{sec:preliminaries}, we construct the equivariant ansatz of the convolutional filters for $\frak{G}_{p4m}$. To start with, we can easily find out that the architecture of QCNN respects the equivariance with respect to $V'_r$ as we repeat the same gate with the same parameter on qubit $i$ and $i+n$ if we have an even $n$. 

The ansatz symmetrization for the other symmetries requires more insights. We will consider the generator gateset, which only consists of Pauli strings up to 2-body local operation : 
\begin{equation}
    G  = \{X, Y, Z, Y_1Y_2, Z_1Z_2\}.
\end{equation}
For a single qubit gate, it is trivial to notice that only Pauli $X$ gates commute with $V_{x}, V_{y}$, while for $k$-qubit gates for $k >1$, we need to explore two cases separately.

\begin{figure}[h]
\centering
    \begin{subfigure}{0.45\textwidth}
        \includegraphics[width = \textwidth, trim = {2.5cm, 9.7cm, 5.5cm, 2cm}]{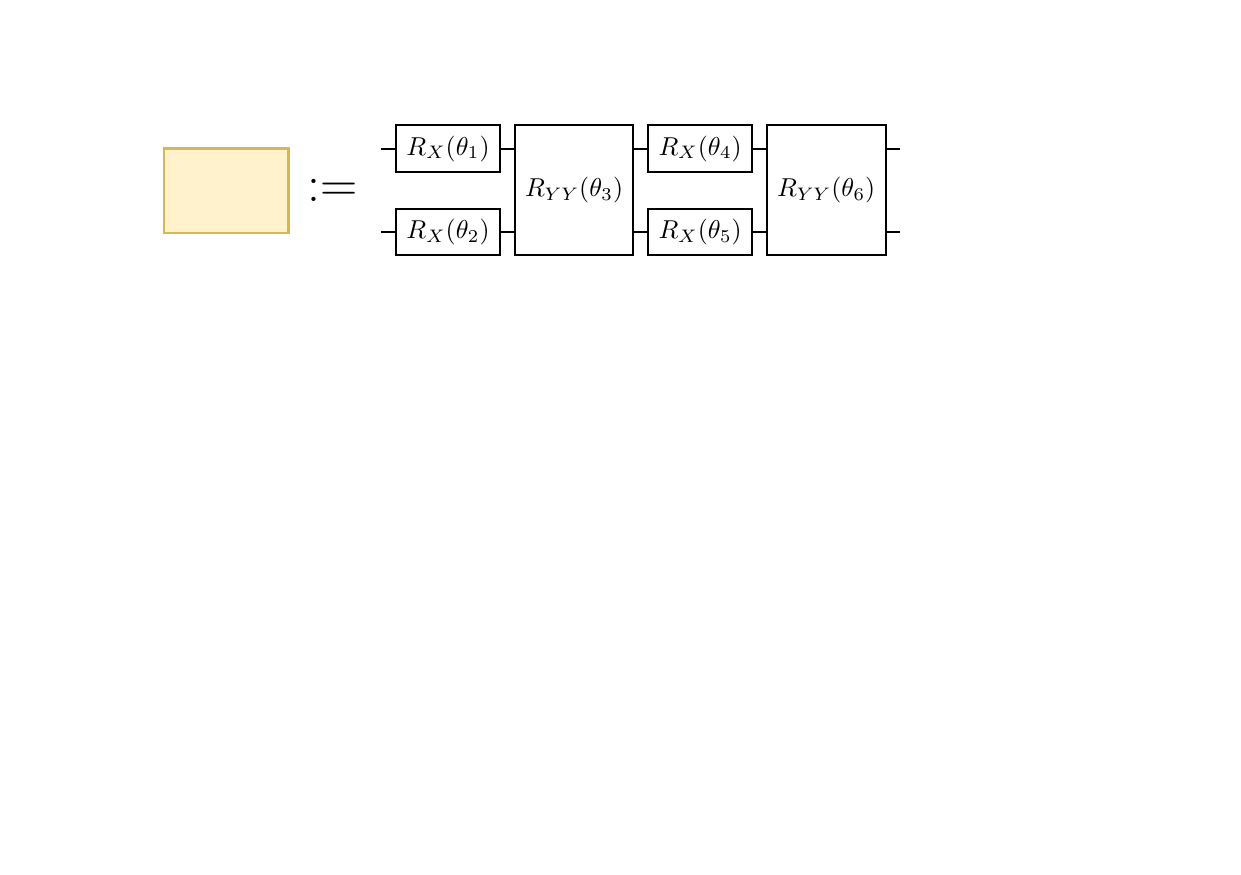}
    \end{subfigure}
    \begin{subfigure}{0.3\textwidth}
        \includegraphics[width = \textwidth, trim = {2.5cm, 3cm, 3cm, 2cm}]{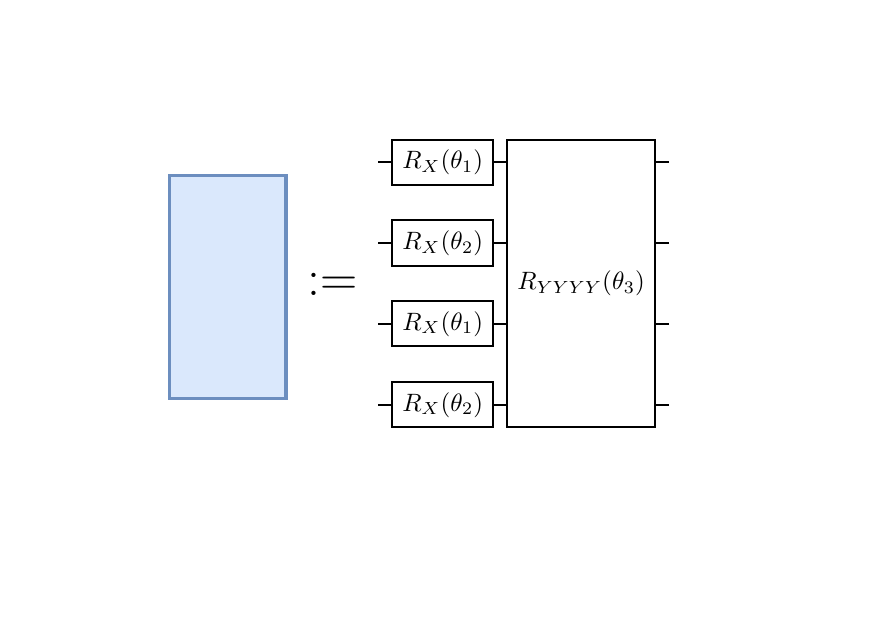}
    \end{subfigure}
\caption{Parameterized quantum circuits ansatz, $U_2$ (yellow rectangle) and $U_4$ (blue rectangle), used the convolutional filters equivariant with respect to $p4m$ symmetry group.}
\label{fig:circuit}
\end{figure}

\begin{enumerate}
    \item \textbf{$G$ constrained to $q_{1:n}$ OR $q_{n+1:2n}$} \\
    Considering only 2-body quantum gates, finding $U \in \frak{comm}(X_{1:n}, X_{n:2n})$ can be simplified into finding $U \in \frak{comm}(X_1X_2)$. We can easily find out that both $Y_1Y_2$ and $Z_1Z_2$ commute with $X_1X_2$ applied on two qubits. Indeed, using the Twirling method, we have : 
    \begin{align}
        \mathcal{T}_{X_1X_2}(Y_1Y_2) & = \frac{1}{2}\Big(Y_1Y_2 + (X_1X_2)^\dagger(Y_1Y_2)(X_1X_2)\Big) \nonumber \\
        & = \frac{1}{2}(Y_1Y_2 + X_1Y_1X_1X_2Y_2X_2)  \nonumber \\
        & =  \frac{1}{2}(Y_1Y_2 + (-Y_1)(-Y_2)).  \nonumber \\
        & = Y_1Y_2. 
    \end{align}
    Similarly, we can show that $Z_1Z_2$ is the equivariant operator with respect to $X_1X_2$. Thus, we obtain the equivariant generator gateset : 
    \begin{equation}
        G_{s,1} = \{Y_1Y_2, Z_1Z_2\}. 
    \end{equation}
    \item \textbf{$G$ applied on both $q_{1:n}$ AND $q_{n+1:2n}$} \\
    Unlike the first case, where the Pauli $X$ gate in $V_x$ and $V_y$ acts equally on two qubits with $X\otimes X$, the weight of Pauli gates is unbiased in this case. We can easily notice that $G_{s,1}$ acting on $q_{n}$ and $q_{n+1}$ do not commute with $V_x$ and $V_y$, as : 
    \begin{equation}
        [X_1\otimes\mathbb{I}_2, Y_1Y_2] = -[Y_1Y_2, X_1\otimes\mathbb{I}_2]. 
    \end{equation}
    Indeed, in order to construct an equivariant operator, we need an even number of Pauli Y or Pauli Z gates applied on both $q_{1\colon n}$ and $q_{n+1\colon2n}$~\cite{nguyen2022theory}. Therefore, the smallest equivariant quantum gates are : 
    \begin{equation}
        G_{s,2} = \{P_\sigma P_\sigma P_{\sigma'} P_{\sigma'} | P_{\sigma, \sigma'} \in \{X, Y, Z\} \} 
    \end{equation}
\end{enumerate}
By exponentiating the equivariant generators found above, we can construct the convolutional filter ansatz, which is equivariant with respect to $V_{p4m}$. 
\figref{fig:circuit} summarizes the equivariant two-qubit convolutional filter ansatz, $U_2$ and the four-qubit ansatz, $U_4$. 
Using the $U_2$ and $U_4$, we propose two QCNN models, Equivariant QCNN (EquivQCNN) and Approximately Equivariant QCNN (Appr-EquivQCNN), as shown in \figref{fig:QCNN}.
In both cases, we first apply the two-qubit convolutional filters on $q_{1:n}$, and $q_{n+1:2n}$ separately, without connecting them, which can be considered as the preliminary scanning phase. Then, in EquivQCNN, $U_4$ ansatz is used as the convolutional filter and connects $q_{1:n}$, and $q_{n+1:2n}$, leading the fully equivariant model. On the other hand, in Appr-EquivQCNN, $U_2$ ansatz is repeated for the learning layers acting on $q_{n}$, and $q_{n+1}$. We add a limited noise to the equivariant model to increase the expressibility by slightly breaking the symmetry. With this noise, we aim to find a crossing point between the expressibility and the equivariance so that it is expressible enough to learn the training samples but, at the same time, not excessively expressible to generalize.

\subsection{Approximately Invariant Measurement}
\begin{figure}[h]
    \centering
    \includegraphics[width = 0.48\textwidth, trim = {2.5cm, 5cm, 1.7cm, 2cm} ]{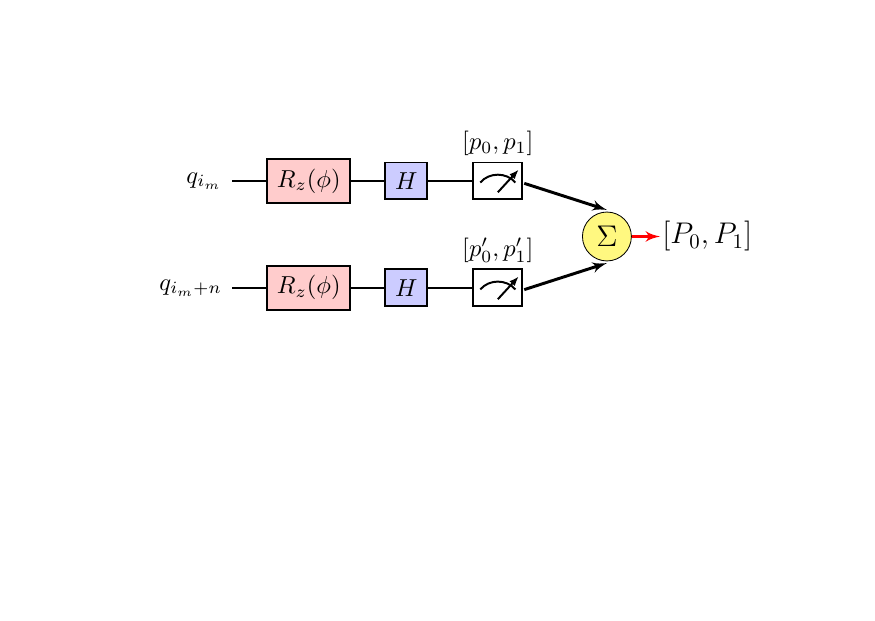}
    \caption{Approximately Invariant Measurement process. We apply $R_z(\phi)$ and $H$ gate on two qubits, $q_{i_m} \in q_{1:n}$ and $q_{i_m+n} \in q_{n+1:2n}$, and measure the probability distribution on each qubits separately. The final label is computed by summing up the two distributions and taking its half.}
    \label{fig:measurement}
\end{figure}
In this section, we propose an \textit{Approximately Invariant Measurement}, with the detailed process summarized in \figref{fig:measurement}. For simplicity, we will consider the binary classification case.

Let us call $q_{i_m} \in q_{1:n}$ and $q_{i_m + n} \in q_{n+1:2n}$ with $i_m \in [1,..,n]$ the qubits which are not traced out in EquivQCNN and measured at the end of the circuit. First, we apply an $R_z(\phi)$ and a Hadamard gate on both $q_{i_m}$ and $q_{i_m + n}$, with $\phi$ also a trained parameter. Then, we measure the probability distribution of state $\ket{0}$ and $\ket{1}$ on each of the qubits separately, obtaining $[p_0, p_1]$ and $[p'_0, p'_1]$ respectively.
As the other qubits are traced out, only the two-qubit state is left at the end of the quantum circuit. Denoting the final quantum state on the qubit $q_{i_m}$ and $q_{i_m + 1}$ as $\ket{\psi_f} = r_0e^{i\theta_0}\ket{00} + r_1e^{i\theta_1}\ket{01} + r_2e^{i\theta_2}\ket{10} + r_3e^{i\theta_3}\ket{11}$, the proposed measurement on $q_{i_m}$ returns the probability distribution : 

\begin{align}
    p_0 = & \frac{1}{2}\big[r_0^2 + r_1^2 + 2r_0r_1\cos(2\phi - \theta_0 + \theta_1) \nonumber \\&  + r_2^2 + r_3^2 + 2r_2r_3\cos(2\phi - \theta_2 + \theta_3)\big],
\end{align}
\begin{align}
    p_1 = & \frac{1}{2}\big[r_0^2 + r_1^2 - 2r_0r_1\cos(2\phi - \theta_0 + \theta_1) \nonumber \\&  + r_2^2 + r_3^2 - 2r_2r_3\cos(2\phi - \theta_2 + \theta_3)\big].
\end{align}

First of all, we can easily notice that the measurement is invariant with respect to $V'_{r}$ as we are summing up the final measurement on $q_{i_m}$ and $q_{i_m + n}$.  
Now, let us prove that it is equivariant with a certain error rate $\epsilon$.  
Similarly, let us call $[p^x_0, p^x_1]$ the final probability of the image reflected with respect to the $x$-axis for the final state $V_{x}\ket{\psi_f} = r_1e^{i\theta_1}\ket{00} +  r_0e^{i\theta_0}\ket{01}  + r_3e^{i\theta_3}\ket{10} + r_2e^{i\theta_2}\ket{11}$ with a bit flip on qubit $q_{i_m}$. By performing the same computation, we can compute the difference between $p_0$ and $p^x_0$ : 
\begin{align}
    p_0 - p^x_0 & = r_0r_1\big[\cos(2\phi  -  \theta_0 + \theta_1) - \cos(2\phi + \theta_0 - \theta_1)\big] \nonumber\\ & + r_2r_3\big[\cos(2\phi  -  \theta_2 + \theta_3) - \cos(2\phi + \theta_2 - \theta_3)\big]  \nonumber\\ 
    & = \sin(2\phi)\big[r_0r_1\sin(\theta_1 - \theta_0) + r_2r_3\sin(\theta_3 - \theta_2)\big] \nonumber \\  
    & \le \frac{1}{2}\epsilon\big[\sin(\theta_2 - \theta_1) + \sin(\theta_3 - \theta_2)] ,  
\end{align}
with $\epsilon = \sin(2\phi)$. The last inequality uses $\max r_0r_1 + r_2r_3 = \frac{1}{2}$ while taking into account the fact that $r_0^2 + r_1^2 +  r_2^2 + r_3^2= 1$. This proves that with $\phi \approx 0$ or $\phi \approx \frac{\pi}{2}$, we can say that the measurement is \textit{approximately invariant} with respect to $V_{x}$, and also $V_{y}$ by using the same justification. The presence of the $R_z$ gate loosens the constraint imposed by the symmetry and adds an extra degree of freedom to the training. By trading off the \textit{full} invariance and expressibility, we allow exploring larger search space for better performance. 
 
We can generalize this measurement for $L$-class classification by measuring $\log_2 L$ qubits at $q_{1:n}$ and $q_{n+1:2n}$ qubits separately and summing them up. This way of measurement corresponds to the \textit{Softmax} activation function at the end of the neural network. Thus, we use the binary cross entropy to calculate the training loss, 
\begin{equation}
    \mathcal{L}_{\bm{\theta}}(\mathbf{x}) = - \sum_{i = 1}^{L} \ell_i \log p_i(\bm{\theta}; \mathbf{x})  , 
\end{equation}
where $\mathbf{\ell} = [\ell_1, \ell_1,..., \ell_{L}]$ with $\ell_i \in \{0,1\}$ is the one-hot encoded target label. 
The state with the highest probability will correspond to the class with which the input image is associated. 

For the following, we will consider two different types of measurements : 
\begin{enumerate}
    \item $M_1$ : $\phi$ is constrained to zero, $\phi = 0$, 
    \item $M_2$ : $\phi$ is updated during the training, $\phi \ne 0$. 
\end{enumerate}

\section{\label{sec:result}Result}
\begin{figure}[h]
\centering
    \begin{subfigure}{0.4\textwidth }
        \includegraphics[width = \textwidth]{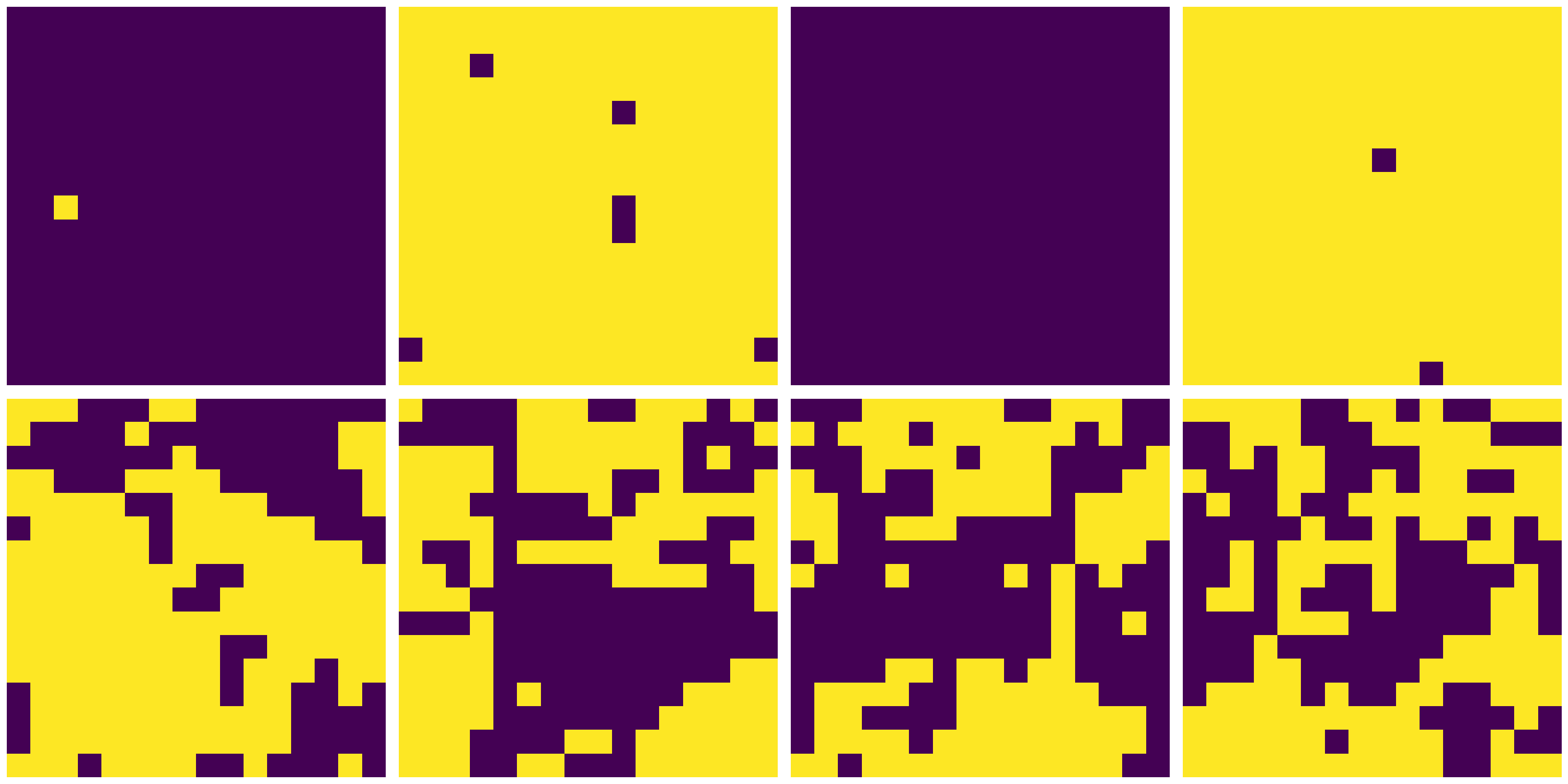}
        \caption{Ising model}
    \end{subfigure}
    \begin{subfigure}{0.4\textwidth }
        \includegraphics[width = \textwidth]{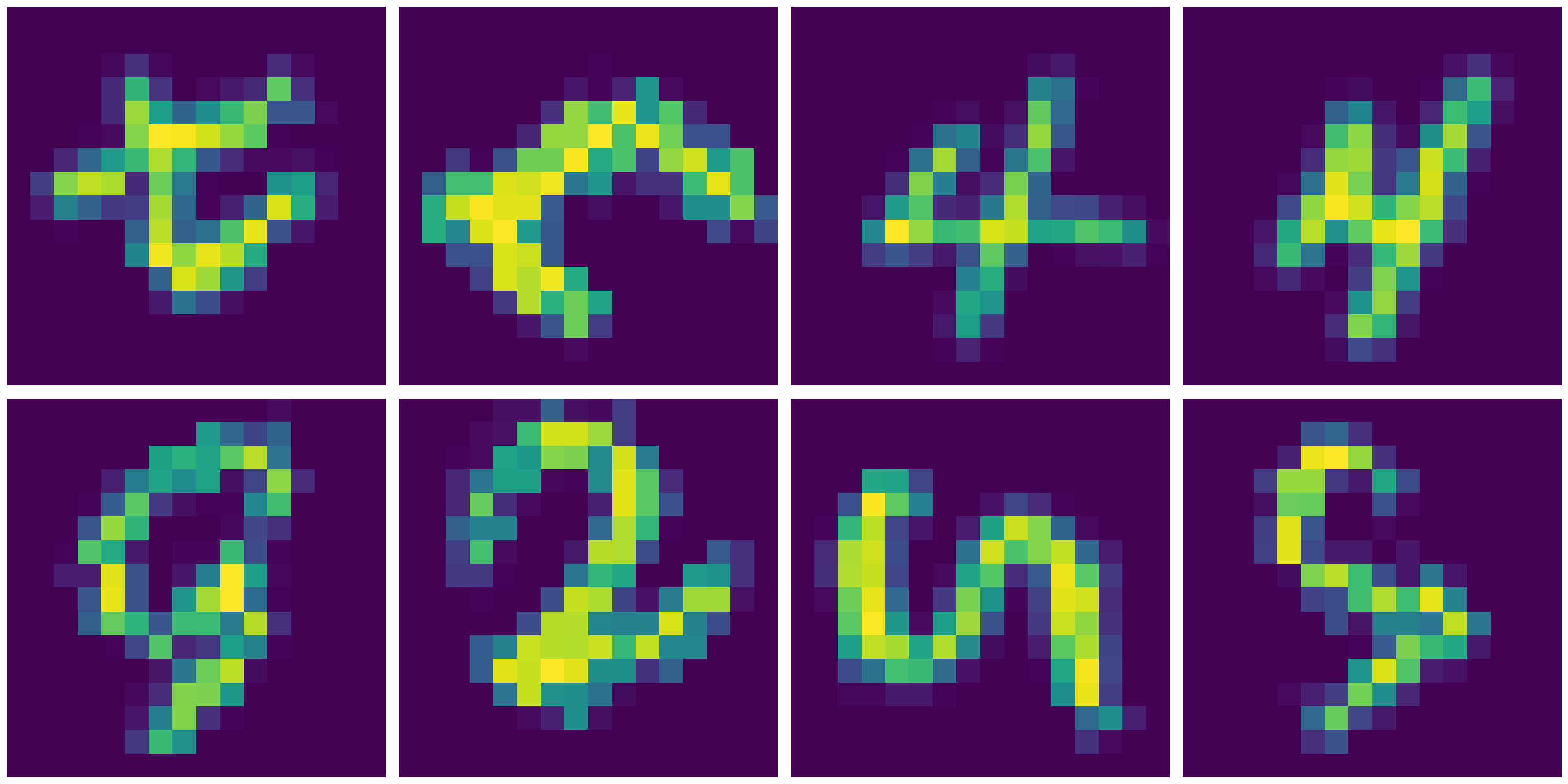}
        \caption{Extended MNIST}
    \end{subfigure}
\caption{Examples of the Ising model and the extended MNIST image samples with size $16\times16$ used for binary classification. Each row corresponds to each class. }
\label{fig:samples}
\end{figure}
In this section, we present our preliminary results of the EquivQCNN training for binary image classification applied to two different image datasets, shown on \figref{fig:samples}. The first dataset contains the spin distribution of the 2D lattice Ising model with $16\times 16$ interacting spins, simulated using Metropolis Monte Carlo with the Hamiltonian~\cite{Onsager1944Ising}: 
\begin{equation}
    H = -J\sum_{\langle ij\rangle} s_is_j, 
\end{equation}
where $s_i\in\{-1, 1\}$ corresponds to the spin on site $i$, $J$ the interaction between two spins and $\langle ij\rangle$ the pairs of the nearest neighbours. At low temperature $T$, the spins are ordered, pointing all in the same direction, and as the temperature increases, we reach the critical temperature $T_c$ where we observe the phase transition from an ordered phase to a disordered phase. By taking the periodic boundary condition, the Ising model dataset is reflectional and rotational symmetric by construction. We aim to classify the order phase from the disordered phase using EquivQCNN. 

The second dataset is the extended MNIST dataset, which also includes randomly reflected and rotated handwritten digit images. In this paper, we present the results for the classification of digits $4$ and $5$, downsampled into $16\times 16$ pixels.

We compare the performance of EquivQCNN with a non-equivariant QCNN with a similar number of parameters, using a convolutional filter that generates an arbitrary two-qubit $SO_4$ state~\cite{Hur2022QCNN}. For all the models, the initial parameters are sampled randomly from a uniform distribution, $[-0.1, 0.1]$. The parameters are updated with ADAM optimizers, using the learning rate of $0.01$, $\beta_1 = 0.5$ and $\beta_2 = 0.999$. 

\begin{table}[h]
    \centering
    \begin{tabular}{c||c|c||c|c}
          & \multicolumn{2}{c||}{Ising} & \multicolumn{2}{c}{MNIST}  \\ 
          \hline
          $n_{samples} $& $40$&  $10240$ &  $40$&  $10240$ \\ 
         \hline
         \hline
         Non-Equiv. & $77.6\pm0.1$& $83.0\pm2.0$ & $66.7\pm2.1$ & $72.9\pm0.5$  \\
         Equiv. & $74.2\pm0.2$ & $75.8\pm0.3$ & $\mathbf{77.5\pm1.0}$ & $74.5\pm 4.7$  \\ 
         Appr-Eq. 1&$84.8\pm2.2$ & $85.4\pm2.0$ & $52.9\pm 0.1$ & $72.7\pm2.4$ \\ 
         Appr-Eq. 2&$\mathbf{86.4\pm3.4}$ &$\mathbf{89.3\pm2.9}$ & $69.7\pm2.9$ & $\mathbf{76.2\pm1.8}$ 
    \end{tabular}
    \caption{The test accuracy at the end of the QCNN training for Ising and extended MNIST with $n_{samples} = 40$ and $10240$ training samples (best result in bold). }
    \label{tab:result1}
\end{table}

\begin{figure}[h]
\centering
    \begin{subfigure}{0.45\textwidth}
        \includegraphics[width = \textwidth]{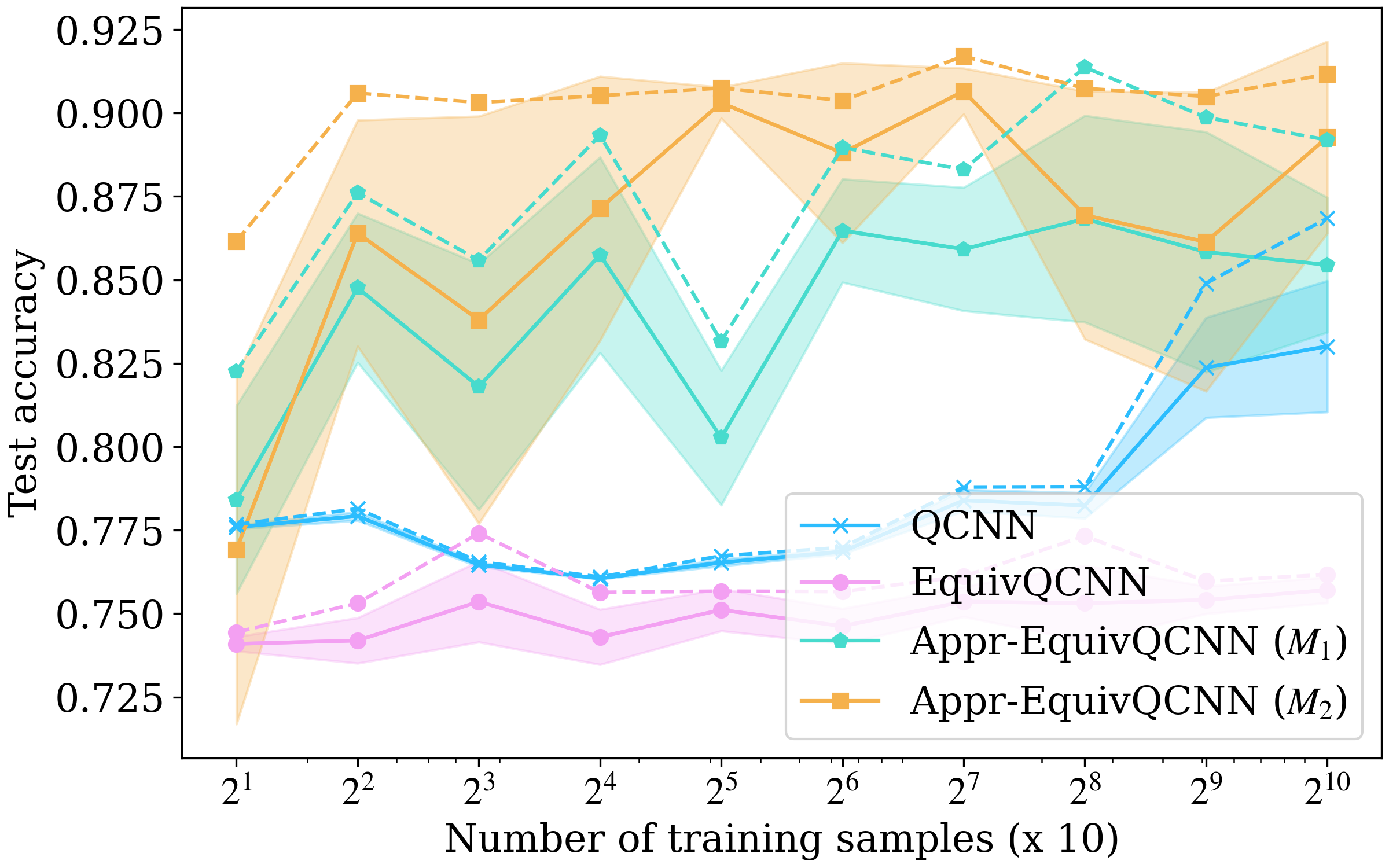}
        \caption{Ising}
    \end{subfigure}
    \begin{subfigure}{0.45\textwidth}
        \includegraphics[width = \textwidth]{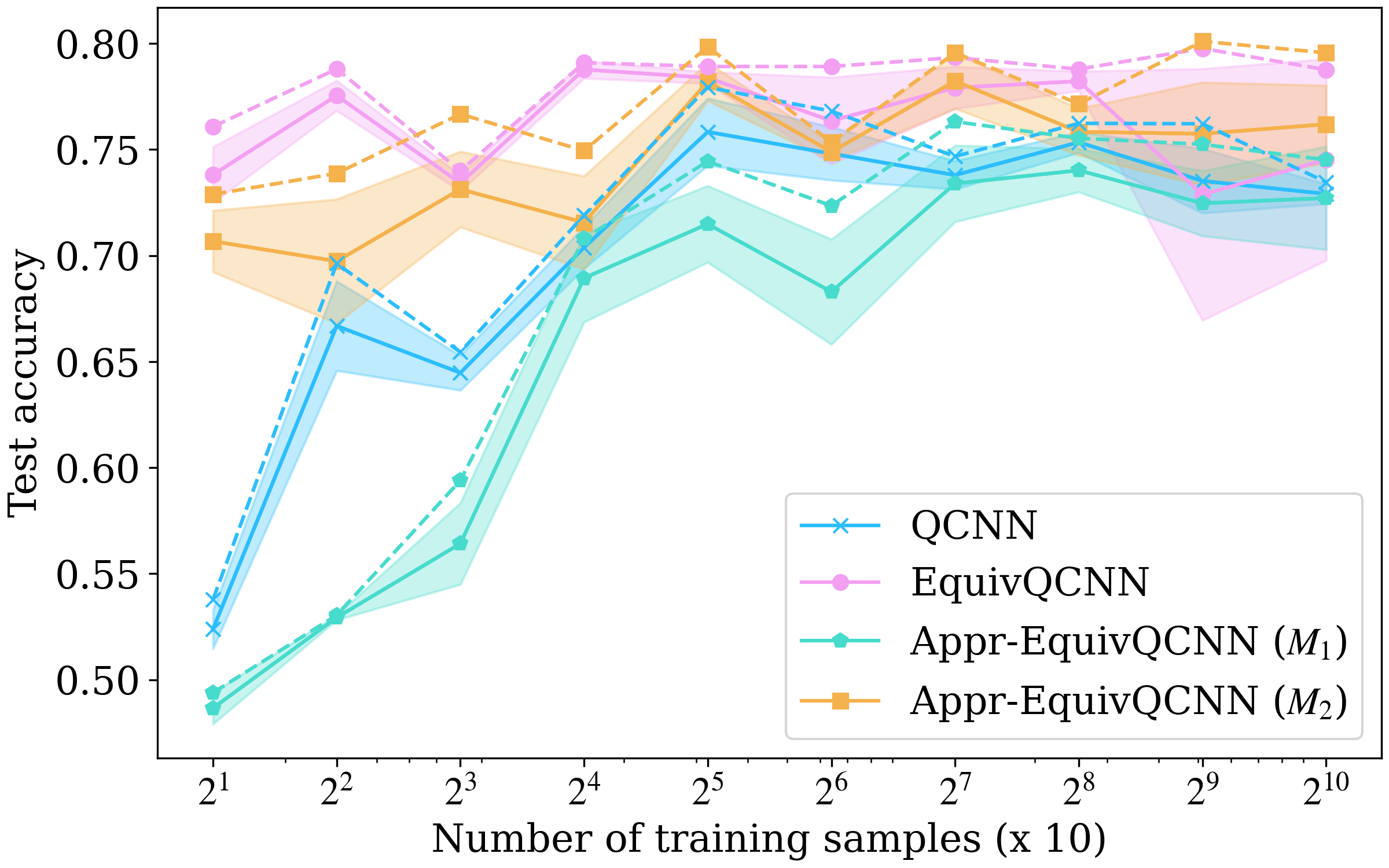}
        \caption{Extended MNIST}
    \end{subfigure}
\caption{The test accuracy obtained at the end of QCNN training for Ising and extended MNIST dataset with different training sample numbers. The solid line corresponds to the average over the five runs, and the dashed line the best one among them. The test accuracy for 
EquivQCNN and ApprEquivQCNN with $M_2$ are always higher than the non-equivariant QCNN, proving their generalization power.}
\label{fig:result2}
\end{figure}

To prove the generalization power of the EquivQCNN, we train the models for different training set size, $N_s = 2^{i}\cdot 10$ for $i = 1,...,10$ with the batch size $N_{bs} = 2^i$ to maintain the same number of updates in each epoch. 
\tabref{tab:result1} and \figref{fig:result2} summarize the test accuracy obtained at the end of the training with different QCNN architectures. Note that the number of samples in the test set is always the same regardless of the number of training samples. 

We can observe that EquivQCNN and Appr-EquivQCNN with $M_2$ measurement have higher test accuracy for all $N_s$, especially for small $N_s$. In particular, in the case of the Ising model, Appr-EquivQCNN with $M_2$ outperforms the non-equivariant model only using 256 times less number of training samples. This certainly proves that the equivariance helps to improve the generalization power as expected. 

One interesting point is that EquivQCNN gives the best result for MNIST, while Appr-EquivQCNN with $M_2$ measurement outperforms EquivQCNN for the Ising model. This difference might be explained by the fact that the Ising model is subject to a stricter symmetry by its construction, compared to the extended MNIST, where the symmetry is artificially created by random reflection and rotation. Thus, injecting noise into the model with Appr-EquivQCNN helps training for the Ising dataset, which is not the case for MNIST.

\section{\label{sec:conclusion}Conclusion}
In this paper, we introduced the Equivariant QCNN for the planar wallpaper symmetry group $p4m$, including reflection and $90^\circ$ rotation in image classification. Furthermore, our study suggests the possibility of injection of noises into the GQML model in order to find the best crossing point between expressibility and equivariance. The proposed models are tested for two different datasets, the Ising model and the extended MNIST dataset, and compared with the non-equivariant model. The results obtained clearly proved that the EquivQCNN outperforms the non-equivariant one in terms of generalization power, especially with a small training set size. Previous studies on QML have already proven that it has a high generalization power with a small training set size~\cite{Caro2022generalization}. This work demonstrated that we can further improve the generalization thanks to the induced bias added by the geometric prior to the dataset.  

For our future research, we plan to compare the EquivQCNN with the problem-agnostic model, not only in terms of test accuracy but also in other factors, such as local effective dimension, overparameterization, barren plateaus, etc. Ultimately, we extend the test to a more realistic use case with a larger image size where symmetry is an essential component for the training, such as Earth Observation images, and show the practical advantage of EquivQCNN. 
\section*{Acknowledgement}
This work was carried out as part of the quantum computing for earth observation (QC4EO) initiative of ESA $\Phi$-lab, partially funded under contract 4000135723/21/I-DT-lr, in the FutureEO programme. MG and SF are supported by CERN through the CERN QTI.

\bibliographystyle{unsrt}
\bibliography{biblio}

\end{document}